\documentclass[aps,pra,twocolumn,superscriptaddress,amssymb,showpacs]{revtex4}
\usepackage{graphicx}
\usepackage{epstopdf}

\begin{document}

\begin{abstract}
{We consider a system consisting of a large individual quantum dot with
excitonic resonance coupled to a single mode photonic cavity in the nonlinear
regime when exciton- exciton interaction becomes important. We show that in the
presence of time-modulated external coherent pumping the system demonstrates
essentially non classical behavior reflected in sub-Poissonian statistics of
exciton- and photon-modes and the Wigner functions with negative values in
phase-space for time-intervals exceeding the characteristic time of dissipative
processes, $t\gg\gamma^{-1}$. It is shown that these results are cardinally
different from the analogous results in the regime of the monomode continues-wave (cw) excitation.}
\end{abstract}


\title{Quantum statistics in time-modulated  exciton-photon system}

\author{G.Yu. Kryuchkyan}
\affiliation{Institute for Physical Researches, National Academy of Sciences, Ashtarak-2, 0203, Ashtarak, Armenia}
\affiliation{Yerevan State University, Centre of Quantum Technologies and New Materials, Alex Manoogian 1, 0025, Yerevan, Armenia}

\author{A.R. Shahinyan}
\affiliation{Institute for Physical Researches, National Academy of Sciences, Ashtarak-2, 0203, Ashtarak, Armenia}

\author{I.A. Shelykh}
\affiliation{Science Institute, University of Iceland, Dunhagi-3, IS-107, Reykjavik, Iceland}
\affiliation{Division of Physics and Applied Physics, Nanyang Technological University 637371, Singapore}
\affiliation{ITMO University, St. Petersburg 197101, Russia}

\maketitle

\section{Introduction}

When light interacts with matter one should discriminate between two
qualitatively different regimes, namely weak and strong coupling. For the first
of them the spectrum of the material system remains unchanged, and light- matter
interaction results in single acts of the emission and absorption of the
photons. On the contrary, in the regime of strong coupling light and matter can
not be anymore treated independently from each other, and hybrid half- light
half matter modes appear in the system. Those modes are known as polaritons.
Nowadays the variety of polaritonic systems is impressive and include such
widely studied cases as surface plasmon polaritons \cite{PlasmonPolarion} and
cavity polaritons \cite{KavokinBook}. In the latter case photonic mode is
coupled with excitonic transition in the quantum well, quantum wire or
individual quantum dot (QD).

Coupling between QD and photonic mode of zero dimensional cavity lies in heart
of such rapidly developing branch of science as cavity quantum electrodynamics
(cQED). The problem is important not only because of the fundamental aspects
brought forward by the interaction of material systems with photons
\cite{Mabuchi02},  but also because of the potential application of cQED to
quantum information processing \cite{Imamoglu99,Bennett00,Chang14}.  From the
point of view of experimental realization, excitons in individual QD can be
brought to strong coupling with confined electromagnetic mode provided by a
pillar (etched planar cavity) \cite{Reithmaier04},  the defect of a photonic
crystal \cite{Yoshie04} or the whispering gallery mode of a microdisk
\cite{Peter05,Kaliteevski07}, among others. In the regime of weak excitation
such structures have demonstrated the Rabi doublet in their optical spectra,
which is characteristic of the mode anticrossing that marks the overcome of
dissipation by the coherent exciton-photon interaction.

These achievements open the way to new research area, namely investigation of the pure quantum effects originating from strong exciton-photon
coupling\cite{Khitrova06}.  Although the system exhibits strong
coupling, it is usually not known in which quantum state it is actually
realized. To be useful for quantum information applications, one should be able to manipulate not just mean number of the particles in the system, but have a tool to monitor and control their statistics. In this context, the possibility to create the states different from essentially classical coherent and thermal states is highly desirable. If the energy of the system
scales linearly with the number of particles, it is essentially classical, \cite{Zhu90} and adding or removing a single
particle will not change its behaviour.  Therefore, the analysis of the nonlinear effects is highly desirable.

With QDs in microcavities, two types of strong nonlinearities are expected,
both associated with the excitons \cite{Laussy05}. The first one
comes from Pauli exclusion, that arises from the
fermionic character of the particles forming an exciton. It becomes extremely
important for the case of the small size QDs where, similar to individual
atoms, the excitation of more then one exciton becomes impossible. Pauli
blocking leads to the radical transformation of the spectrum of the system,
which changes from the Rabi doublet to Mollow triplet when the intensity of the
external pump is increased \cite{Muller07,Flagg09,Ates09,DelValle10,Laussy08}.

The second is Coulomb repulsion between the excitons, again arising from their composite nature. This mechanism is important for the case of the excitons in the QWs and large QDs. In the former case it leads to the blueshift of the polaritonic modes increasing with the intensity of the external pump- the effect which can be satisfactory described within the frameworks of the mean- field approximation. The case of an individual large QD inside the cavity is, however, more tricky. Coulomb repulsion between the excitons in this configuration leads to the emergence of rich multiplet structure in the emission spectrum \cite{LaussyShelykh}, which reveals the pure quantum nature of light- matter coupling.

In the present paper we analyze further the essentially quantum effects arising from nonlinearities in coupled QD- cavity system. The novelty of our scheme stems from the idea that quantum systems can display qualitatively new forms of behavior when driven by fast time-periodic modulations. Particularly, the application of a sequence of tailored pulses as well as time-modulated cw field leads to  improving the degree of quantum effects in open cavity nonlinear systems and onset of qualitatively new quantum effects. This approach was recently exploited for the formation of a high degree continuous-variable entanglement in the nondegenerate
optical parametric oscillator \cite{adam, adam2}, for generation of Fock states
in Kerr nonlinear resonator (KNR) driven by a sequence of Gaussian pulses
\cite{gev1, gev2} and for demonstration of multiphoton blockades in pulsed
regimes of  dissipative KNR beyond stationary limits \cite{far,gor}. It has
also demonstrated that amplitude modulation can improve the performance of
single photon sources based on quantum dot \cite{ser}. The idea to enrich
quantum physical systems by designing a time modulation has been explored in
several other fields of research including periodically driven
nonlinear oscillator \cite{dyk} and periodically driven quantum matter \cite{gold}.

Here we focus on consideration of cavity modes in the regimes of strong
exciton-photon coupling and strong exciton-exciton interaction with respect to
the rates of damping of the photonic- and exciton-modes. In these regimes the
transition frequencies between energy levels of the systems without any
interaction with external field are differently spaced in the quantum regime.
Thus, we enable spectroscopic identification and selective excitation of
transitions between combined exciton-photon number states. This consideration
provides the framework of master equation and the numerical method of
quantum trajectories on the base of excitation numbers, $Q$ Mandel parameter
or the second-order correlation function. We focus also  on the Wigner
functions of exciton and photonic modes that allow phase-space monitoring
exciton-photon coupling in quantum treatment.

We show that in the presence of time-modulation the ensemble-averaged mean
photon numbers, the populations of photon-number and exciton-number states, and
the Wigner functions are nonstationary and exhibit a periodic time dependent
behavior, i.e. repeat the periodicity of the pump laser in an over transient
regime.  We demonstrate the possibility of the observation of purely quantum
effects, namely sub-Poissonian statistics of exciton- and photon-modes and the
Wigner functions with negative values in phase-space for time-intervals
exceeding the characteristic time of dissipative processes, $t\gg\gamma^{-1}$.
It is shown that these results are cardinally different from the analogous
results in cw regime of excitation. Beside this we  investigate temperature
noisy effects for a cavity at finite temperatures. It leads to applications in
simulating of more realistic exciton-photon systems as well as to study of
unusual quantum phenomena connecting quantum engineering and temperature.

The paper is organized as follows. In Sec. II, we present the effective
Hamiltonian for the periodically driven polariton system and describe the
physical quantities of interest. In Sec. III we study time-evolution of the
mean photon numbers as well as the exciton numbers, quantum statistics of
excitonic and photonic modes on base of the $Q$ Mandel parameter and the
second-order correlation function for zero-delay time.  We analyze also the
distributions of photon-number states and  the phase-space  properties of
photon-and exciton-modes on base of the Wigner functions. The effects coming
from a thermal reservoir is also briefly analysed.  We summarize our results in
Sec. IV.

\section{Coupled exciton-photon system}

The system consists of coupled  fundamental photonic dot  and  exciton modes driven by cw field with mean frequency $\omega$ and time-modulated amplitude. The Hamiltonian of driven photon-exciton system in the rotating wave approximation (RWA) reads:

\begin{eqnarray}
H= \Delta_{ph} a^{\dagger}a +  \Delta_{ex} b^{\dagger}b +  \chi {b^{\dagger}}^{2} b^{2} +
 g(b a^{\dagger} + b^{\dagger} a) + \nonumber \\
+(\Omega_1 + \Omega_2 exp(-i\delta t))a + H.c ,\label{hamiltonianL}
\end{eqnarray}

where $a^{+}$, $a$ are creation and annihilation operators of the photon mode,
$b^{+}$, $b$ are creation and annihilation operators for the exciton mode, $g$
is the exciton-photon coupling constant, $\chi$ is the strength of
exciton-exciton interaction, $\Delta_{ph}=\omega_{ph} -\omega$,
$\Delta_{ex}=\omega_{ex} -\omega$ are detunings between mean frequency of the
driving field and the frequencies of the photonic and exciton modes. $\Omega_1$
and $\Omega_2$ are the components of complex amplitude of the driven field and
$\delta$ is the frequency of the modulation. The case $\Omega_2=0$  describes
the exciton-photon cavity driven by cw monochromatic field treated within RWA.
Such situation can be realized also if photon-exciton system is driven by two
fields with different frequencies. In this case,  the Hamiltonian of the system
in RWA is reduced to Eq. \ref{hamiltonianL} with $\delta$ being the difference
between frequencies of the driving fields.

In realistic systems one should necessarily take into account the dissipation,
because the modes suffer from losses due to partial transmission of light
through the mirrors of the photonic cavity, non-radiative decay of excitons and
decoherence. We consider these effects by assuming that the interaction of
driven photon-exciton system with heat reservoir gives rise to the damping
rates of modes $\gamma_{a}$ and   $\gamma_{b}$.
 We trace out the reservoir degrees of freedom in the Born-Markov limit
assuming that system and environment are uncorrelated at initial time t = 0.
This procedure leads to the master equation for the reduced density matrix  in
the Lindblad form. The master equation within the framework of the
rotating-wave approximation, in the interaction
picture corresponding to the transformation $\rho  \rightarrow   e^{-i\omega
a^{+} a t}\rho e^{i\omega  a^{+} a t}$ reads as

\begin{equation}
\frac{d \rho}{dt} =-i[H, \rho] +
\sum_{i=1,2,3,4}\left( L_{i}\rho
L_{i}^{\dagger}-\frac{1}{2}L_{i}^{\dagger}L_{i}\rho-\frac{1}{2}\rho L_{i}^{\dagger}
L_{i}\right)
\label{master},
\end{equation}
where $L_{1}=\sqrt{(n_{th}+1)\gamma}a$, $L_{2}=\sqrt{n_{th}\gamma}a^+$,  $L_{3}=\sqrt{(n_{th}+1)\gamma}b$ and  $L_{4}=\sqrt{n_{th}\gamma}b^+$are the
Lindblad operators, $\gamma$ is a dissipation rate, and $n_{th}$ denotes the mean
number of quanta of a heat bath.  Here, for simplicity we assume that the  decay rates and  the mean
numbers of quanta of a heat bath are equal for the both modes, $\gamma_{a}=\gamma_{b}=\gamma$.

We analyze the  mean values of excitation numbers
as well as we probe the strength of quantum fluctuations of exciton and photonic modes via the Mandel factor and the  normally ordered
second- order  correlation functions of the photon
numbers and excitation numbers. Beside this we monitor phase properties of both modes by the Wigner function. These  quantities of interest are calculated
by using the reduced density operators of the photons $\rho_{a}(t)$ and of the excitons $\rho_{b}(t)$.  These operators are
constructed from the full density operator of the system $\rho(t)$  by
tracing out the excitonic or photonic modes respectively,
\begin{eqnarray}
\rho_{a}(t)=Tr_{b}(\rho),\\
\rho_{b}(t)=Tr_{a}(\rho).
\end{eqnarray}
For the system under time modulated external pumping the ensemble-averaged mean oscillatory
excitation numbers as well as the other physical quantities exhibit a periodic time-dependent behavior after initial transient time interval.
Using the master equation above we calculate the time-evolution of the $Q$ Mandel factor for the photonic and exciton modes. For photonic mode it is defined as
\begin{equation}
Q(t) = \frac{\langle(\Delta n(t))^2\rangle - \langle n(t) \rangle}{\langle n(t) \rangle}
\end{equation}
where $\langle(\Delta n(t))^2\rangle = \langle(a^{\dagger}a )^2\rangle - (\langle a^{\dagger}a \rangle)^2$ describes the deviation of the excitation number uncertainty from the Poissonian variance, $\langle(\Delta n)^2\rangle=\langle n \rangle$.  The case $Q = 0$ corresponds to Poissonian statistics. If $Q > 0$, the statistics is super-Poissonian, if $Q < 0$ it is sub-Poissonian, and analogous for the exciton mode.

The Mandel factor is connected with the normalized second-order correlation function for zero delay time $ g^{(2)}$ defined (for the photonic mode) as:
\begin{equation}
g^{(2)}(t)=\frac{\langle a^{\dagger}(t)a^{\dagger}(t)a(t)a(t)\rangle}{(\langle a^{\dagger}(t)a(t)\rangle)^2}.
\end{equation}
For a short counting time-intervals the approximate relation between these quantities reads as
 $\langle (\Delta n)^2\rangle = \langle n\rangle + \langle n\rangle ^2(g^{(2)}-1)$.
Thus, the condition $g^{(2)}<1$ corresponds to the sub-Poissonian statistics, $\langle (\Delta n)^2\rangle < \langle n\rangle$.

We analyze the master equation numerically using well known quantum state
diffusion method  \cite{dqs}. According to this method, the reduced density
operator is calculated as the ensemble mean over the stochastic states
describing evolution along a quantum trajectory.

It should be mentioned that exciton-photon system is actually presented as the
model of driving  coupled oscillators with anharmonic term. Such model  has
been focus of considerable
attention and this interest is justified by many applications in different contexts. Particularly, in field of quantum devices for a few-photon level one application concerns to realization of strong photon blockade in photonic molecules with modest Kerr-nonlinearity of the photon using two coupled photonic cavities \cite{liew, bamba, maj}.

\section{Improvement of quantum effects by periodic time-modulation}

In our further consideration we concentrate on analysis of coupled exciton-photon in quantum regimes at
low level of quanta. This analysis is performed  by using the following set of dimensionless parameters $\Delta/\gamma
$, $\chi/\gamma$, $g/\gamma$, as well as parameters of amplitude modulation
$\Omega_1/\gamma$,  $\Omega_2/\gamma$ and the frequency of modulation $\delta$.

Let us discuss the operational regimes in more details. If  energy levels of
the coupled exciton- photon states are well resolved we can consider near to
the resonant selective transitions between lower photon-exciton states. In this
way, the detunings play an important role when identifying the spectral lines,
thus we arrange the detunings of modes to reach a qualitative quantum effects.
We estimate the detuning by using the results on the structure of energy levels
of  a quantum dot in a microcavity in the nonlinear regimes  Ref.
\cite{LaussyShelykh}. In this paper the optical spectrum at resonance
transition  was studied  by diagonalizing the Hamiltonian \ref{hamiltonianL} without the
driving term. In this case the total number of excitations is conserved, and
eigenstates of the total particle number $a^{\dagger} a+b^{\dagger} b$ with
eigenvalue $m$ can be used as the $m$th manifold of the system. Particularly,
two-photon excitation of the system with $m=2$ leads to the eigenvectors that
invoves triplet states $|2,0\rangle$, $|1,1\rangle$, $|0,2\rangle$ with
photonic and exciton number states. If we neglect the exciton-exciton
interaction, the  energy levels form the Rabi triplet $E_{+}=2\omega_{ph} +
2g$, $E_{0}=2\omega_{ph}$, $E_{-}=2\omega_{ph} -2g$. In this way, two-photon
resonant frequencies leading to exsitation of the level $E_{-}$ equals to
$2\omega_2=E_{-}-E_{g}$ , where $E_{g}$ is the level of ground state.
Therefore,
$\omega_2=\omega_{ph} - g$ and hence the detunings $\Delta_{ex}=\omega_{ex}
-\omega_{2}$, $\Delta_{ph}=\omega_{ph} -\omega_{2}$ are
$\Delta_{ph}=\Delta_{ex}=g$. In this approach the  effects of exciton-exciton
interaction can be included by using numerical calculations  Ref.
\cite{LaussyShelykh}. For strong driving field the detunings can be also
shifted due to Stark effects. Thus, for the concrete calculations, later we use
the approximate value of detunings corresponding to the parameters $g/\gamma$
and $\chi/\gamma$. Note, that such analysis seems to be rather qualitative than
quantitative but allows us to estimate the values of detunings. Indeed,
considering strong analysis of the selective excitation of cavity modes we need
in calculations of the other resonant  frequencies corresponding the large
numbers of quanta, $m>2$. Below, for concrete calculations we use the values
$\Delta_{ph}/\gamma = \Delta_{ex}/\gamma = 7.12$ that  are estimated for the
parameter $g/\gamma$ = 7.

\begin{figure}
\includegraphics[width=7.6cm]{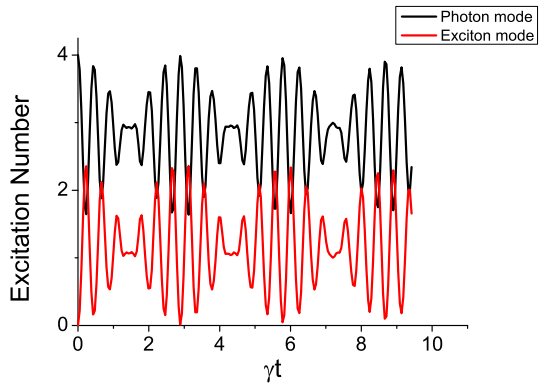}
\caption{Time-evolution of the  mean excitation numbers of photonic and excitonic modes for the nondissipative case, without any driving.  The parameters are as follows: $\Delta_{ph}/\gamma=\Delta_{ex}/\gamma=7.12$, $g/\gamma$ = 5, $\chi/\gamma$ = 4.}
\label{nonDis}
\end{figure}

 For the illustrative purposes let us first consider the case of the
non-dissipative system. The typical results for the time-evolution of the mean
excitation numbers of both photon-mode and exciton-mode due to the coupling
between modes and the exciton-exciton nonlinearity are shown at the Figs.
\ref{nonDis}. We assume that at $t=0$ the exciton mode is in vacuum state $
\langle n(0) \rangle_{b}=0$ while the four-photon number state is injected into
the resonator, $ \langle n(0) \rangle_{a}=4$. As we see, without any driving
the total number of quanta is conserved and the dynamics of excitation numbers
display oscillatory behaviour, where are observed the Rabi transitions between
the modes with collapses and revivals effects.

The collapses and revivals are well known phenomena in quantum
optics, particularly, in the context of the Jaynes-Cummings model (see, for example \cite{rev}) and for ion-trap system \cite{monroe}, \cite{jakob}. Here, we obtain interesting
collapses and revival effects  in the simplest model as a manifestation of  nonlinear exciton-exciton interaction that leads to spaced energies of exciton number states. Note, also that in this case conservation of total particle numbers  lead to anti-phase dynamics of mode that is violated in the presence of driving.

\begin{figure}
\includegraphics[width=8.6cm]{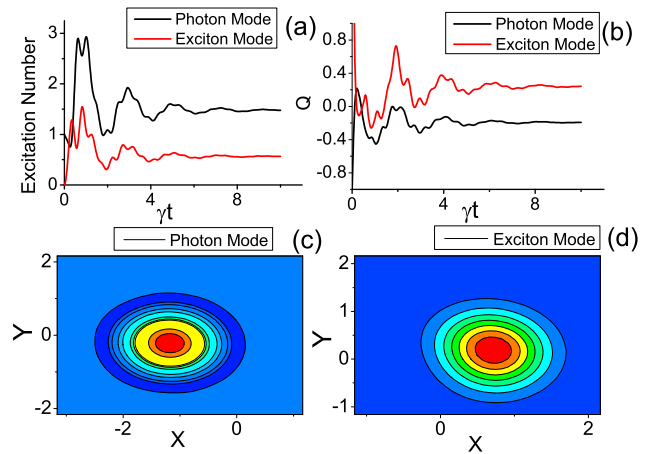}
\caption{The results for the case of single mode  driving, $\Omega_2=0$. (a) temporal dependence of the mean excitation numbers of modes; (b) temporal dependence of the Mandel parameters for excitonic and photonic modes.  The contour plots of the Wigner functions of  photonic mode (c) and excitonic mode (d) for time intervals corresponding to the maximal values of excitation numbers exceeding the characteristic decoherence time. The parameters are as follows: $\Delta_{ph}/\gamma = 7.12$, $\Delta_{ex}/\gamma = 7.12$,
$g/\gamma$ = 5, $\chi/\gamma$ = 1, $\Omega_1/\gamma$ = 5, $\Omega_2/\gamma$ = 0, $\delta/\gamma=0$.}
\label{linStat}
\end{figure}

Now we turn to the realistic case where dissipation and decoherence are taken
into account considering  pure quantum effects for zero temperature cavity.  We
start from the case of $\Omega_2$ = 0, which corresponds to the single mode cw
excitation, and the initial state corresponding to no excitations in the
cavity. The typical results for the excitation numbers, the Mandel parameters
and the contour plots of the Wigner functions of two modes are depicted in
Fig. \ref{linStat}. As we see,   the mean excitation numbers and the Mandel
parameter show Rabi-like oscillations for short time-intervals in the transient
regime and reach the equilibrium in the steady-state regime. From  Fig.
\ref{linStat}(a) we conclude that  for used parameters the system is operated
in strong quantum regime at level of small excitation numbers.
 In stationary, over-transition regime  the exciton Mandel parameter  $Q_{b} > 0$, while  $Q_{a} =-0.2$ in steady state limit, i.e. photonic mode displays sub-Poissonian statistics.  The Wigner functions of both mode are positive in all phase space but contours of the Wigner function for photonic mode have slightly squeezed form corresponding to small sub-Poissonian statistics.

\begin{figure}
\includegraphics[width=8.6cm]{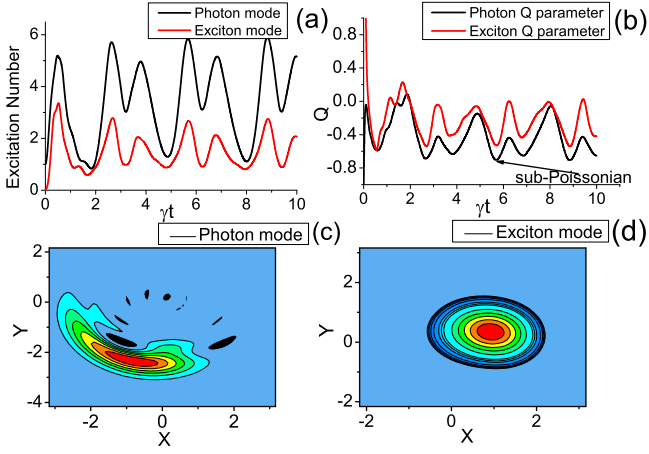}
\caption{The results for the case of periodically driven photon-exciton system.  (a) temporal dependence of the  mean excitation numbers for the excitonic and photonic modes; (b) temporal dependence of the Mandel parameters for both modes; (c) the Wigner function of the photonic mode corresponding to the maximal occupancy of the photonic mode; (d) the Wigner function of the photonic mode corresponding to the minimal occupancy of the photonic mode. The parameters are as follows: $\Delta_{ph}/\gamma = 7.12$, $\Delta_{ex}/\gamma = 7.12$,
$g/\gamma$ = 5, $\chi/\gamma$ = 1, $\Omega_1/\gamma$ = 5, $\Omega_2/\gamma$ = 5, $\delta/\gamma=2$.}
\label{linMod}
\end{figure}

\begin{figure}
\includegraphics[width=8.6cm]{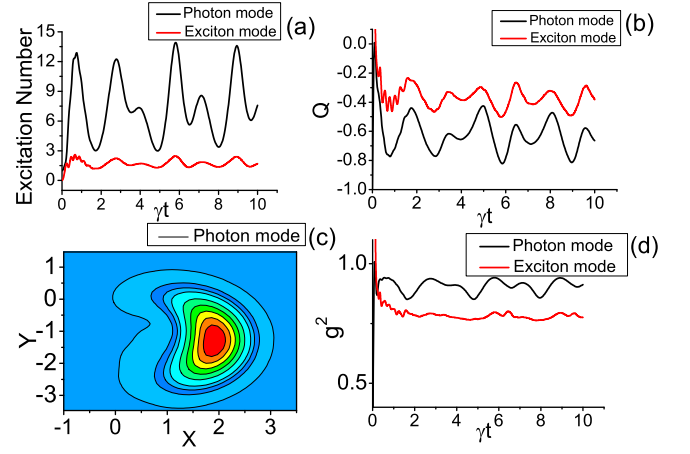}
\caption{The results for the case of periodically driven photon-exciton system in deeply sub- Poissonian regime. (a) time-dependence of
the  mean excitation numbers; (b) time-dependence of the Mandel parameters for
excitonic and photonic modes; (c) the contour plots of the Wigner function
corresponding to the maximal occupancy of the photonic mode; (d) time-dependent normalized second-order correlation functions for photonic- and exciton-modes. The parameters are
as follows: $\Delta_{ph}/\gamma = 7.12$, $\Delta_{ex}/\gamma = 7.12$,
$\chi/\gamma$ = 3, $g/\gamma$ = 7, $\Omega_1/\gamma$ = 5, $\Omega_2/\gamma$ =
5, $\delta/\gamma=2$.}
 \label{bestMandel}
\end{figure}

The typical results for the case of time-modulated excitation with
$\Omega_2\neq0$ are presented in the Figs. \ref{linMod}, \ref{bestMandel} and
\ref{interC} for various parameters of the exciton-photon system. As one can
see at the Figs. \ref{linMod}(a) and \ref{linMod} (b),  for the coupling
constants: $\chi/\gamma$ = 1 and  $g/\gamma = 5$, the mean excitation numbers
and the Mandel parameters repeat the periodicity of the pump laser in an over
transient regime. It is remarkable that time-modulation of modes leads to
formation of highly sub-Poissonian statistics  compared to the case of
monochromatic driving (see, results depicted in Fig. \ref{linStat}). Indeed,
in this case for the photonic mode the values of Mandel parameter can reach the
values of $Q_{a} \approx-0.65$ and for the exciton mode $Q_{b} \approx-0.45$.
If one compare this numbers with those obtained earlier for a single mode cw
pumping, one can make a conclusion that quantum effects in the behavior of
exciton-photon system becomes more pronounced for the case of the time-
modulated pumping. Note, that most negative values of the Mandel parameters are
realized for the time-intervals corresponding to the maximal values of the
occupations of the modes. The quantum effects are thus more essential for the
comparatively large number of the quanta. The contour plots of the Wigner
functions for maximal and minimum values of photonic mode are presented in
Figs. \ref{linMod}(c),(d), respectively.  Note, that for the parameters
considered when the photon excitation number corresponds to its maximum, the
corresponding Wigner function has a region in phase-space where it is
negatively defined (these regions are shown in black). On the contrary, for the
times corresponding to the minimal photon occupancies the Wigner function is
almost Gaussian and Mandel parameters are very close to zero. This indicates
that in these moments the state of the system is very close to coherent and its
behavior is essentially classical.

Increasing of the coupling constant of the exciton- photon coupling and the
strength of  excition- exciton nonlinear interaction makes quantum effects more
pronounced, as it is shown in Fig. \ref{bestMandel} for both excitonic and
photonic modes and the parameters: $\chi/\gamma$ = 3, $g/\gamma = 7$. Note,
that for these parameters the detunings  $\Delta_{ph}/\gamma = 7.12$,
$\Delta_{ex}/\gamma = 7.12$ approximately correspond to two-photon selective
excitation of the level  $E_{-} $ from vacuum state as it has been shown above.

 The  time-evolution of averaged excitation numbers are depicted in Fig.
\ref{bestMandel}(a). Comparing these results with analogous ones shown in
Fig. \ref{linMod}(a) we conclude, that increasing of the parameters
$\chi/\gamma$ and  $g/\gamma$ and leaving the other parameters without change,
leads to increasing
of the level of  photon excitation numbers and decreasing the excitation
numbers of exciton mode.
Considering quantum statistics of modes  (see, Fig. \ref{bestMandel}(b)) we
conclude that this regime displays a deeply sub-Poissonian statistics with
Mandel parameter achieving the values of $Q_{a} \approx-0.8$ for photonic mode
and  $Q_{b} \approx-0.5$  for excitonic mode in their minima. It is also
interesting to present results on quantum statistics of exciton-photon system
within the framework of the normalized second-order correlation function by using
the above formulas. The results for nonstationary correlation functions for
photonic and excitonic modes versus dimensionless time intervals  are depicted
in  Fig. \ref{bestMandel}(d). As we see, in this regime  the results display
photon anti-bunching as well as  the exciton anti-bunching for all time-intervals.
However, the anti-bunching for exciton mode is slightly stronger than for the
photonic mode.
 Beside this, the correlation function of exciton-mode shows  monotonous
time-behaviour, with $g^{(2)} \approx0.76$,  while analogous result for photonic
mode is non-monotonous with $g^{(2)} \approx0.82$ at its minima at definite
time intervals corresponding to the minimal values of the mean photon number.
Note, for the comparison that for $|2\rangle$ pure Fock state the normalized
second-order photon
correlation function equals to 0.5. Thus, the obtained results reflect
excitations of states with high-order $m>2$th manifolds  and  probably
describe nonclassically  of photon-exciton states for moderate number of
quanta. A more complete description of the system  can be obtained by calculation
of the Wigner function in phase space. In this way, the contour plots  of the
Wigner function for photonic mode
depicted in Fig. \ref{bestMandel}(c) displays the typical form corresponding to light with  sub-Poissonian statistics.

\begin{figure}
\includegraphics[width=8.6cm]{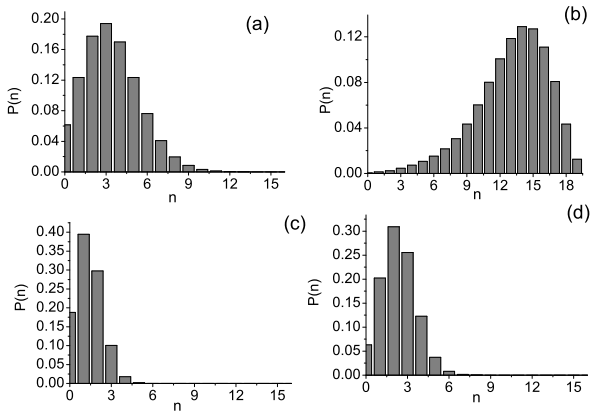}
\caption{ The photon number distributions at the minimum level of quanta (a) and  at the maximum level of quanta (b).  The  exciton number distributions at the minimum level of quanta (c) and  at the maximum level of quanta (d). The parameters are
as follows: $\Delta_{ph}/\gamma = 7.12$, $\Delta_{ex}/\gamma = 7.12$,
$\chi/\gamma$ = 3, $g/\gamma$ = 7, $\Omega_1/\gamma$ = 5, $\Omega_2/\gamma$ =
5, $\delta/\gamma=2$.}
\label{Fig5}
\end{figure}

Additionally, the probability distributions of photon-number and exciton-number
states for the definite time-intervals are presented in Fig. \ref{Fig5}. These
results shows that non-selective  excitations of both photonic and excitonic
modes are realized for the parameters used. Indeed, the distributions of photon
numbers are relatively large and centred around the maximal and minimal values
of mean photon numbers (see Figs. \ref{bestMandel} (a), (b)).
Nevertheless, the distributions
considerable differs from the Poissonian distribution. The distributions of
exciton numbers correspond to excitations of mode at level of a few quanta in
accordance with the results of mean exciton numbers.

\begin{figure}
\includegraphics[width=8.6cm]{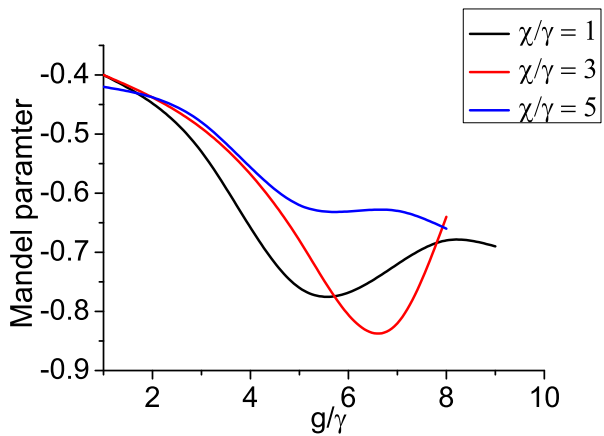}
\caption{ The dependence of the minimal value of the Mandel parameter on the exciton- photon coupling strength for several values of the exciton- exciton interaction: (a) $\chi/\gamma$ = 1; (b)$\chi/\gamma$ = 3; (c)$\chi/\gamma$ = 5. The other parameters are taken as: $\Delta_{ph}/\gamma = 7.12$, $\Delta_{ex}/\gamma = 7.12$, $\Omega_1/\gamma$ = 5, $\Omega_2/\gamma$ = 5, $\delta/\gamma=2$.}
\label{interC}
\end{figure}

It is interesting to analyse the minimal values of the Mandel factor in its
time-evolution  for various coupling constants of the exciton-photon coupling
and the strengths of excition- exciton nonlinear interaction. The results for
photonic mode  in dependence from the coupling constant $g/\gamma$ are
presented in Fig. \ref{interC} for three values of the parameter $\chi/\gamma$
and the fixed value of the amplitude of driving field.  One can clearly see
that for the definite parameters of driving time-dependent field,
exciton-exciton interaction and the detunings there exist an optimal value of
the ratio $g/\gamma$ where the quantum effects in the behavior of the system
becomes most pronounced. These results indicate the regime presented in Fig.
\ref{bestMandel} as most preferable for production of strong sub-Poissonian
statistics. It should be mentioned that the  parameters of the Gaussian pulses
in above consideration are  free parameters and they might be chosen in order
to find optimal regimes producing   high degree of  sub-Poissonian statistics
or photon antibunching.

At the end of this section we turn to thermal effects considering briefly the
interaction of the exciton-photon system with the reservoir at finite
temperatures.  We investigate how the temperature affects the Mandel
factor and the correlation function of photon mode in comparison to the case
of zero-temperature reservoir.  The effects coming from thermal reservoirs are
interesting for performing more realistic approach to generate nonclassical
states in exciton-photon systems  and for study of phenomena connecting quantum
engineering and temperature.

\begin{figure}
\includegraphics[width=8.6cm]{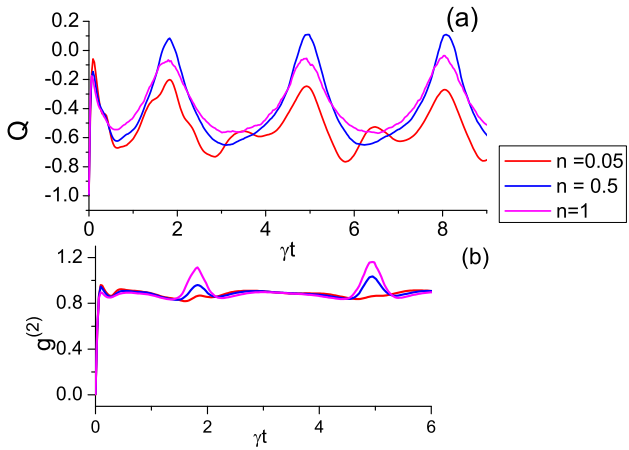}
\caption{Time evolution of the  Mandel factor for three values of thermal photon numbers $n_{th}$ (a).  The autocorrelation function versus time-intervals for the several numbers of thermal photons  (b).
 The parameters are as follows:
$\Delta_{ph}/\gamma = 7.12$, $\Delta_{ex}/\gamma = 7.12$,
$\chi/\gamma$ = 1, $g/\gamma$ = 5, $\Omega_1/\gamma$ = 5, $\Omega_2/\gamma$ =
5, $\delta/\gamma=2$.}
\label{01}
\end{figure}

The results for thermal photons  in the range $n_{th}=0.01-1$ are shown in
Figs. \ref{01}(a),(b). Taking into account that $T=\hbar\omega_{th}/k_B$, for
the frequencies of thermal photons $\omega_{th}=10 GHz$ the temperature in
this range  corresponds to
$10^{-1}K - 10K$.
 In order to illustrate the difference between the case of zero-temperature resonator we assume here the parameters as in the previous case depicted in  Fig. \ref{linMod}.
As our calculation shows, the maximal values of mean photon numbers and the mean exciton numbers are
approximately the same as in the case of pure resonator,
while quantum statistics of oscillatory modes is changed due to the thermal noise. The light inside the cavity remains sub-Poissonian and anti-bunched for all time intervals,  if  $n_{th}=0.05$. In this case the minimal values of Q-parameter and the correlation function remains nearly to the case of vacuum reservoir. As expected, the further increasing of the temperature leads to decreasing of quantum effects.

\section{conclusions}

In conclusion, we considered the quantum effects in a system consisting of an individual large quantum dot containing interacting excitons and single mode photonic cavity. We have shown that the nature of the external driving can substantially change the behavior of the system. Namely quantum effects become more pronounced in the case of the time modulated two mode driving as compared to the simple single mode external pump.

We thank Prof. Yu. G. Rubo, Dr. A. Sheremet and Prof. O.V. Kibis for the discussions we had on the subject. The work was supported by FP7 ITN "NOTEDEV". A.R. Shahinyan thanks the University of Iceland for hospitality and G. Yu. Kryuchkyan acknowledge support from the Armenian State Committee of Science, the Project No.13-1C031.

\end{document}